\RequirePackage{ifpdf}
\ifpdf 
\documentclass[pdftex]{sigma}
\else
\documentclass{sigma}
\fi

\newcommand\caA{{\mathcal A}}
\newcommand\caB{{\mathcal B}}
\newcommand\caF{{\mathcal F}}
\newcommand\caG{{\mathcal G}}

\newcommand\caL{{\mathcal L}}

\newcommand\caM{{\mathcal M_\theta}}
\newcommand\caS{{\mathcal S}}

\newcommand\caZ{{\mathcal Z}}
\newcommand\wx{{\widetilde x}}
\newcommand\gone{{ \mathchoice {1\mskip-4mu\mathrm{l} } {1\mskip-4mu\mathrm{l} }{1\mskip-4.5mu\mathrm{l} } {1\mskip-5mu\mathrm{l}} }}
\newcommand\gR{{\mathbb R}}

\newcommand\gC{{\mathbb C}}
\newcommand\gP{{\mathbb P}}
\newcommand\gN{{\mathbb N}}
\newcommand\gZ{{\mathbb Z}}

\newcommand\algzero{{\mathsf 0}}
\newcommand\algA{{\mathbf A}}
\newcommand\algrA{{\mathbf A}^\bullet}

\newcommand\kg{{\mathfrak g}}

\newcommand\kh{{\mathfrak h}}

\newcommand\kX{{\mathfrak X}}
\newcommand\kY{{\mathfrak Y}}

\newcommand\ad{{\text{\textup{ad}}}}
\newcommand\Ad{{\text{\textup{Ad}}}}

\newcommand\fois{\mathord{\cdot}}
\DeclareMathOperator{\tr}{Tr} 

\newcommand\Der{{\text{\textup{Der}}}}
\newcommand\Int{{\text{\textup{Int}}}}

\newcommand\dd{{\text{\textup{d}}}}
\newcommand\tF{{\text{\textup{F}}}}

\begin{document}

\allowdisplaybreaks

\renewcommand{\thefootnote}{$\star$}

\renewcommand{\PaperNumber}{048}

\FirstPageHeading

\ShortArticleName{On the Origin of the Harmonic Term in Noncommutative Quantum Field Theory}

\ArticleName{On the Origin of the Harmonic Term\\ in Noncommutative Quantum Field Theory\footnote{This paper is a
contribution to the Special Issue ``Noncommutative Spaces and Fields''. The
full collection is available at
\href{http://www.emis.de/journals/SIGMA/noncommutative.html}{http://www.emis.de/journals/SIGMA/noncommutative.html}}}

\Author{Axel de GOURSAC}

\AuthorNameForHeading{A. de Goursac}

\Address{D\'epartement de Math\'ematiques, Universit\'e Catholique de Louvain, \\
Chemin du Cyclotron, 2, 1348 Louvain-la-Neuve, Belgium}
\Email{\href{mailto:axelmg@melix.net}{axelmg@melix.net}}

\ArticleDates{Received March 30, 2010, in f\/inal form June 01, 2010;  Published online June 09, 2010}

\Abstract{The harmonic term in the scalar f\/ield theory on the Moyal space removes the UV-IR mixing, so that the theory is renormalizable to all orders. In this paper, we review the three principal interpretations of this harmonic term: the Langmann--Szabo duality, the superalgebraic approach and the noncommutative scalar curvature interpretation. Then, we show some deep relationship between these interpretations.}

\Keywords{noncommutative QFT; gauge theory; renormalization; Heisenberg algebra}

\Classification{81T13; 81T15; 81T75}

\renewcommand{\thefootnote}{\arabic{footnote}}
\setcounter{footnote}{0}

\section{Introduction}
\label{sec-intro}

Since around f\/ifteen years, the interest for noncommutative geometry has been growing in physics. In many domains of mathematics, like topology, measure theory, geometry, one has noticed that spaces are equivalently described by the commutative algebras of their functions. Then, noncommutative geometry \cite{Connes:1994,Landi:1997} considers noncommutative algebras as corresponding to some ``noncommutative spaces''. A particular type of noncommutative algebras is given by deformation quantization of symplectic spaces. This is the case of the Moyal space \cite{Moyal:1949}, and it can be physically interpreted as the fact that one cannot know simultaneously positions in the space with an arbitrary precision, but at a certain scale (the Planck scale) events are no more localizable \cite{Doplicher:1994tu}. Physical theories on such noncommutative spaces, for instance noncommutative quantum f\/ield theory, are then candidates for new physics beyond the Standard Model of par\-tic\-le physics. Of course, the study of gauge theory on noncommutative spaces is of fundamental importance, as it is the case in the Standard Model.

On the Euclidean Moyal space, the real scalar $\phi^4$ theory is not renormalizable, because of a~new type of divergence called the Ultraviolet-Infrared (UV/IR) mixing \cite{Minwalla:1999px}, and which seems to be generic on several noncommutative spaces \cite{Gayral:2004cs,Gayral:2005af}. Renormalizability is however a~fundamental property for physical f\/ield theories. Recently, a solution has been proposed \cite{Grosse:2004yu} by adding a~harmonic term in the action, so that the theory is renormalizable at all orders. As we will see it in this paper, this model has new interesting f\/low properties, which do not appear in the usual commutative scalar theory. The mathematical understanding of this harmonic term is of f\/irst importance since it may permit to generalize this solution to other noncommutative spaces.

We propose here to review the three principal mathematical interpretations of the harmonic term: the Langmann--Szabo duality \cite{Langmann:2002cc} and its group formulation (in \cite{Bieliavsky:2008qy}), the superalgebraic approach \cite{deGoursac:2008bd}, and the interpretation in terms of a noncommutative scalar curvature \cite{Buric:2009ss,Buric:2010xs}. Then, we show how the Langmann--Szabo duality can be reformulated in the superalgebraic framework at the group level, and we underline the relationship between the three viewpoints.

The paper is organized as follows. We f\/irst present the Moyal space and the UV/IR mixing in Subsections~\ref{subsec-qft-moyal} and~\ref{subsec-qft-uvir}. Then, the solution with harmonic term is exposed with its proper\-ties in Subsection~\ref{subsec-qft-harm}, so as its associated gauge theory, obtained by an ef\/fective action, in Subsection~\ref{subsec-qft-gauge}. The three already quoted possible origins of the harmonic term  are reviewed in Section~\ref{sec-inter}. Finally, we discuss the unif\/ication and the relations between these interpretations in Section~\ref{sec-disc}.

\section{Quantum f\/ield theory on the Moyal space}
\label{sec-qft}

\subsection{Presentation of the Moyal space}
\label{subsec-qft-moyal}

We expose in this subsection the basic def\/initions and properties of the Moyal space \cite{Moyal:1949} for the paper to be self-contained. The Moyal space is a deformation quantization of the space $\gR^D$, for an even dimension~$D$. Let $\Sigma$ be the symplectic form on $\gR^D$ represented in the canonical basis by the matrix:
\begin{gather}
\Sigma=\begin{pmatrix} 0 & -1 & 0 & 0 & \\ 1 & 0 & 0 & 0 & \\ 0 & 0 & 0 & -1 & \ddots \\ 0 & 0 & 1 & 0 &  \\  &  & \ddots &  & \end{pmatrix},\label{eq-qft-sigma}
\end{gather}
 and let $\Theta=\theta\Sigma$, where $\theta$ is a real parameter of the deformation. On $\caS(\gR^D)$, the $\gC$-valued Schwartz functions space on $\gR^D$, is def\/ined the following associative noncommutative product: $\forall\, f,g\in\caS(\gR^D)$, $\forall\,  x\in\gR^D$,
\begin{gather*}
(f\star g)(x)=\frac{1}{\pi^D\theta^D}\int \dd^Dy\,\dd^Dz\ f(x+y) g(x+z)e^{-iy\wedge z},
\end{gather*}
where $y\wedge z=2y_\mu\Theta^{-1}_{\mu\nu}z_\nu$ (we use in this paper the Einstein summation convention). It is called the Moyal product, and turns $\caS(\gR^D)$ into a topological $\ast$-algebra (with the usual involution ${}^\dag$).

Then, this product can be extended on the temperated distributions $\caS'(\gR^D)$ by duality, and one consider $\caM$, the algebra of (left and right) multipliers of $\caS(\gR^D)$ in $\caS'(\gR^D)$, endowed with the Moyal product. $\caM$ is also a topological $\ast$-algebra involving $\caS(\gR^D)$ and polynomial functions for instance. See \cite{GraciaBondia:1987kw,Varilly:1988jk} for more details.

The limit $\theta\to 0$ is called the commutative limit because the Moyal product
\begin{gather*}
(f\star g)(x)|_{\theta=0}=f(x)g(x)
\end{gather*}
is the usual pointwise commutative product in this limit. Moreover, the integral is a trace for the Moyal product since if $f,g\in\caM$ such that $f\star g\in L^1(\gR^D)$, one has
\begin{gather}
\int \dd^Dx\, (f\star g)(x)=\int\dd^Dx\, f(x)g(x).\label{eq-qft-trace}
\end{gather}
The usual derivatives $\partial_\mu$ are inner derivations:
\begin{gather*}
\partial_\mu f=-\frac i2[\wx_\mu,f]_\star,
\end{gather*}
where $\wx_\mu=2\Theta^{-1}_{\mu\nu}x_\nu$, the commutator is $[f,g]_\star=f\star g-g\star f$
and the anticommutator $\{f,g\}_\star=f\star g+g\star f$. Note that the coordinate functions satisfy the following commutation relation:
\begin{gather*}
[x_\mu,x_\nu]_\star=i\Theta_{\mu\nu}.
\end{gather*}
Finally, the Moyal space is a spectral triple of the non-compact type \cite{Gayral:2003dm}, and its spectral distance has been computed in \cite{Cagnache:2009ik}.

\subsection{The UV-IR mixing}
\label{subsec-qft-uvir}

The straightforward generalization of the real $\phi^4$ theory on the Euclidean Moyal space:
\begin{gather*}
S(\phi)=\int \dd^Dx\left(\frac 12\partial_\mu\phi\star\partial_\mu\phi +\frac{m^2}{2}\phi\star\phi +\lambda\phi\star\phi\star\phi\star\phi\right),
\end{gather*}
suf\/fers from a new type of divergence, called the ultraviolet-infrared (UV/IR) mixing \cite{Minwalla:1999px}. Let us here analyze this divergence. Thanks to property \eqref{eq-qft-trace}, the action is given by:
\begin{gather}
S(\phi)=\int \dd^Dx\left(\frac 12(\partial_\mu\phi)^2 +\frac{m^2}{2}\phi^2 +\lambda\phi\star\phi\star\phi\star\phi\right),\label{eq-qft-actinit}
\end{gather}
so that only the interaction is changed by a non-local term. Any vertex of this theory admits only cyclic permutations as symmetries and not all permutations like in the commutative case (for $\theta=0$). In the Fourier space, the Feynman rules are:
\begin{itemize}
\itemsep=0pt
\item $\frac{1}{p^2+m^2}$ for a propagator,
\item $\lambda\,e^{i\frac{\theta^2}{4}(p_1\wedge p_2+p_1\wedge p_3+p_2\wedge p_3)}$ for a vertex,
\end{itemize}
where $(p_1,p_2,p_3,p_4)$ are the incoming impulsions to the vertex, with respect to its cyclic order.

\begin{figure}[htb]
  \centering
  \includegraphics{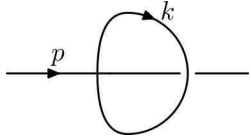}
  \caption{Non-planar tadpole.}
  \label{fig-qft-nonplanartadpole}
\end{figure}

By taking into account the cyclic order of each vertex, the Feynman graphs can be divided into two sectors: planar graphs and non-planar ones (for more details, see \cite{Filk:1996dm}). The non-planar graphs have new IR-divergences, called UV/IR mixing. Let us see such an example of divergence, which is typical of the UV/IR mixing as shown in \cite{Magnen:2008pd}. The computation of the amplitude associated to the non-planar tadpole (see Fig.~\ref{fig-qft-nonplanartadpole}) gives rise to
\begin{gather*}
\caA(p)=\frac{\lambda}{(2\pi)^D}\int \dd^Dk\, \frac{e^{\frac{i\theta^2}{2}k\wedge p}}{k^2+m^2} =\frac{\lambda}{(2\pi)^{\frac D2}}\left(\frac{m^2}{\theta^2p^2}\right)^{\frac{D-2}{4}} K_{\frac D2-1}(m\theta|p|),
\end{gather*}
where $p$ is the external impulsion and $K$ is a modif\/ied Bessel function. This amplitude is f\/inite for a f\/ixed external impulsion $p\neq 0$, but singular by taking the limit $|p|\to 0$. Indeed, in $D=4$ dimensions, $\caA(p)\propto_{p\to 0}\frac{1}{p^2}$.

Then, if this non-planar tadpole diagram is inserted into higher-loop order graphs, this impulsion $p$ is integrated over and can produce an IR-divergence. Since this divergence comes from the noncommutativity of the limit $\Lambda\to\infty$, for an UV regularization $\Lambda$ of the variable $k$, and of the IR-limit $|p|\to0$, it is called UV/IR mixing. And this divergence cannot be renormalized by counterterms of the form of the initial action \eqref{eq-qft-actinit}, so that the theory is not renormalizable. Numerical evidence for the ef\/fects induced by the UV/IR mixing can be found for example in~\cite{Panero:2006cs,Panero:2006bx}.

\subsection{Noncommutative scalar f\/ield theory with harmonic term}
\label{subsec-qft-harm}

The f\/irst solution to this problem of UV/IR mixing in the real scalar f\/ield theory has been proposed by H.~Grosse and R.~Wulkenhaar \cite{Grosse:2004yu} by adding a harmonic term in the action:
\begin{gather}
S(\phi)=\int \dd^Dx\left(\frac 12(\partial_\mu\phi)^2 +\frac{\Omega^2}{2}\wx^2\phi^2 +\frac{m^2}{2}\phi^2 +\lambda\phi\star\phi\star\phi\star\phi\right),\label{eq-qft-actharm}
\end{gather}
where $\Omega$ is a real parameter. The propagator of this theory is changed and given by the Mehler kernel in the position space \cite{Gurau:2005qm}:
\begin{gather}
C(x,y) =\frac{\theta}{4\Omega}\left(\frac{\Omega}{\pi\theta}\right)^{\frac D2}\int_0^\infty  \frac{\dd\alpha}{\sinh^{\frac D2}(\alpha)} e^{-\frac{m^2\alpha}{2\widetilde \Omega}}C(x,y,\alpha),\nonumber\\
C(x,y,\alpha)  = \exp\left(-\frac{\widetilde \Omega}{4}\coth\left(\frac \alpha 2\right)(x-y)^2 -\frac{\widetilde \Omega}{4}\tanh\left(\frac \alpha 2\right)(x+y)^2\right).\label{eq-qft-propag}
\end{gather}
where $\widetilde \Omega=\frac{2\Omega}{\theta}$.

Of course, this theory breaks now the translation invariance (indeed $C(x,y)\neq C(x-y)$). But this lack of translation invariance is responsible of the removing of the UV/IR mixing. The amplitude of the non-planar tadpole (Fig.~\ref{fig-qft-nonplanartadpole}) has the same behavior when $|p|\to 0$ as before (see Subsection~\ref{subsec-qft-uvir}), but the propagators joining these non-planar tadpoles in a higher loop order graph are now of the type \eqref{eq-qft-propag} and remove the IR-divergence thanks to this translational symmetry breaking~\cite{VignesTourneret:2006xa}. And the theory \eqref{eq-qft-actharm} has been shown to be renormalizable to all orders in perturbation in $D=4$ \cite{Grosse:2004yu,Rivasseau:2005bh,Gurau:2005gd,Gurau:2007fy} and superrenormalizable in $D=2$~\cite{Grosse:2003nw}. A parametric representation of this model has been provided in \cite{Gurau:2006yc,Rivasseau:2007qx} and a Connes--Kreimer Hopf algebra encoding its renormalization has also been constructed in~\cite{Tanasa:2007xa,Tanasa:2009hb}.

Let us study some properties of the noncommutative scalar f\/ield theory with harmonic term~\eqref{eq-qft-actharm}.
\begin{itemize}
\itemsep=0pt
\item Concerning the renormalization f\/lows, it has been shown up to three loops that the constant $\Omega$ was running towards a f\/ixed point $\Omega=1$ \cite{Disertori:2006uy}. At this f\/ixed point, the beta function of the coupling constant $\lambda$ vanishes up to irrelevant terms at all orders of perturbation \cite{Disertori:2006nq,Grosse:2004by}, so that the theory \eqref{eq-qft-actharm} does not involve any Landau ghost, contrary to the commutative $\phi^4$ model.
\item Moreover, the vacuum solutions of the theory which respect the same symmetry have been exhibited in every dimensions \cite{deGoursac:2007uv,deGoursac:2009gh}, in view of a possible spontaneous symmetry breaking.
\item Furthermore, even if the deformation quantization, and in particular the choice of the symplectic structure $\Sigma$ with $\Theta=\theta\Sigma$, breaks the rotation group symmetry, it has been shown \cite{deGoursac:2009fm} that the rotational invariance is fully restored at the classical and quantum level (at all orders in perturbation) by considering a family of actions labeled by the symplectic structures of a certain orbit of the rotation group.
\end{itemize}

Note that there are now other renormalizable theories on the Moyal space. For instance, the LSZ-model \cite{Langmann:2003if} and the Gross--Neveu model \cite{VignesTourneret:2006nb}, respectively in the complex scalar case ($D=4$) and in the fermionic case ($D=2$). Another renormalizable real scalar model on the Moyal space has been exhibited \cite{Gurau:2008vd}, in which the non-local IR counterterm $\frac{1}{p^2}\phi^2$ is now included in the classical action. The resulting theory is translation-invariant, but does not possess the properties exposed above for the Grosse--Wulkenhaar model (see also \cite{Geloun:2008hr,Tanasa:2010fk,Magnen:2008pd,Blaschke:2008yj,Blaschke:2009gm}).

\subsection{The associated gauge theory}
\label{subsec-qft-gauge}

Let us f\/irst introduce the noncommutative framework adapted to the $U(1)$-gauge theory on the Moyal space. See \cite{deGoursac:2007gq,Wallet:2007em} for more details. Gauge potentials are real elements $A_\mu\in\caM$ ($A_\mu^\dag=A_\mu$) and the associated covariant derivatives can be expressed as $\forall\phi\in\caM$,
\begin{gather*}
\nabla_\mu\phi=\partial_\mu\phi-iA_\mu\star\phi.
\end{gather*}
In this setting, gauge transformations are determined by unitary elements $g\in\caM$ ($g^\dag\star g=g\star g^\dag=1$), and act on the f\/ields as
\begin{gather*}
\phi^g =g\star\phi,\qquad
A_\mu^g =g\star A_\mu\star g^\dag+ig\star\partial_\mu g^\dag,
\end{gather*}
so that $\nabla_\mu\phi$ transforms as: $(\nabla_\mu\phi)^g=g\star(\nabla_\mu\phi)$. Note that the gauge theory has also an adjoint action on complex scalar f\/ields $\varphi\mapsto g\star\varphi\star g^\dag$, as it will be the case in Subsection~\ref{subsec-inter-super}. Like in the commutative case (but for non-Abelian theories), the curvature of the potential $A_\mu$ takes the form:
\begin{gather*}
F_{\mu\nu}=\partial_\mu A_\nu-\partial_\nu A_\mu-i[A_\mu,A_\nu]_\star,
\end{gather*}
and transforms covariantly under gauge transformations: $F_{\mu\nu}^g=g\star F_{\mu\nu}\star g^\dag$.

A major dif\/ference with the commutative case is the existence of a canonical gauge-invariant connection. By setting $\xi_\mu=-\frac 12\wx_\mu$, it turns out that the gauge potential $A_\mu^{\text{inv}}=\xi_\mu$ def\/ines a~connection invariant under gauge transformations: $(A_\mu^{\text{inv}})^g=g\star A_\mu^{\text{inv}}\star g^\dag+ig\star\partial_\mu g^\dag=A_\mu^{\text{inv}}$. Let us introduce the covariant coordinates $\caA_\mu=A_\mu-A_\mu^{\text{inv}}$, transforming also covariantly: $\caA_\mu^g=g\star\caA_\mu\star g^\dag$.

The natural action for the $U(1)$-gauge theory on the Moyal space is
\begin{gather}
S(A)=\int\dd^Dx\left(\frac 14F_{\mu\nu}\star F_{\mu\nu}\right),\label{eq-qft-actgauge}
\end{gather}
but it suf\/fers also from the UV/IR mixing, which renders its renormalizability quite unlikely~\cite{Matusis:2000jf}. Indeed, by taking into account the ghost contribution, the ``non-planar'' polarization tensor is f\/inite for a f\/ixed external impulsion $p\neq 0$, but singular in $D=4$ by taking the limit $|p|\to 0$:
\begin{gather*}
\Pi^{\text{np}}_{\mu\nu}(p)\propto_{p\to0}\frac{\widetilde p_\mu\widetilde p_\nu}{p^4}.
\end{gather*}
The same phenomenon as in the scalar theory appears (see Subsection~\ref{subsec-qft-uvir}), and one can hope that an analogous solution of the problem of UV/IR mixing can be found.

By investigating what could be the analogue of the harmonic term for gauge theory, a gauge action has been computed from the Grosse--Wulkenhaar model coupled with gauge f\/ields by a~one-loop ef\/fective action \cite{deGoursac:2007gq,Grosse:2007dm}:
\begin{gather}
S(A)=\int\dd^Dx\left(\frac 14F_{\mu\nu}\star F_{\mu\nu}+\frac{\Omega^2}{4}\{\caA_\mu,\caA_\nu\}_\star^2+\kappa \caA_\mu\star\caA_\mu\right),\label{eq-qft-acteff}
\end{gather}
where $\Omega$ and $\kappa$ are new real parameters. This gauge-invariant action is naturally associated to the scalar theory with harmonic term \eqref{eq-qft-actharm} and is therefore a good candidate to renormalizability. Note that the two additional terms are typically noncommutative since they depend on $\caA_\mu$ and on the existence of a canonical connection.

Upon expanding the quadratic term in $A_\mu$ of the action \eqref{eq-qft-acteff}, by using $\caA_\mu=A_\mu+\frac 12\wx_\mu$, one f\/inds:
\begin{gather}
\int\dd^Dx\left(-\frac 12A_\mu\partial^2 A_\mu+\frac{\Omega^2}{2}\wx^2A_\mu A_\mu+\kappa A_\mu A_\mu-\frac 12\big(1-\Omega^2\big)(\partial_\mu A_\mu)^2+\Omega^2(\wx_\mu A_\mu)^2\right).\label{eq-qft-actquadr}
\end{gather}
Except the last two (of\/f-diagonal) terms, which need to be suppressed by an appropriate gauge f\/ixing, this expression is exactly the quadratic operator of the scalar theory \eqref{eq-qft-actharm} applied on the f\/ield $A_\mu$, with a harmonic term and a mass term, and whose inverse is the Mehler kernel, responsible of the removing of the UV/IR mixing in the scalar case. Notice that the ghost sector of a similar gauge model has been studied in~\cite{Blaschke:2007vc,Blaschke:2009aw}.

However, the action involves also a linear part in $A_\mu$:
\begin{gather*}
\int\dd^Dx\left(\frac{\Omega^2}{2}\wx^2\wx_\mu A_\mu+\kappa\wx_\mu A_\mu\right),
\end{gather*}
which means that $A_\mu=0$ is not a solution of the equation of motion. A simple vacuum $A_\mu=A_\mu^{\text{inv}}$ (or equivalently $\caA_\mu=0$) leads to non-dynamical matrix model \cite{deGoursac:2007qi}. The other vacuum solutions have been exhibited in \cite{deGoursac:2008rb}. For a review on noncommutative QED with strong background f\/ields, see \cite{Ilderton:2010rx}.

\section{The possible origins of the harmonic term}
\label{sec-inter}

In this section, we review the dif\/ferent mathematical interpretations of the harmonic term of the model~\eqref{eq-qft-actharm}, which is responsible of the renormalizability and of the special f\/low properties of this noncommutative theory.

\subsection[Langmann-Szabo duality]{Langmann--Szabo duality}
\label{subsec-inter-lsd}

The Grosse--Wulkenhaar model \eqref{eq-qft-actharm} has a special symmetry pointed out by Langmann and Sza\-bo~\cite{Langmann:2002cc}, and called the Langmann--Szabo duality. This symmetry is in fact a (cyclic) symplectic Fourier transformation: for $\phi$ a real scalar f\/ield,
\begin{gather}
\hat\phi(k_a)=\frac{1}{(\pi\theta)^{\frac D2}}\int\dd^Dx\,\phi(x)e^{-i(-1)^ak_a\wedge x},\label{eq-inter-symplfour}
\end{gather}
where $a\in\{1,\dots,4\}$ denotes the position of $\hat\phi(k_a)$ in a product $\hat\phi\star\dots\star\hat\phi$, or equivalently the (cyclic) order of the external impulsion $k_a$ in a vertex. This Fourier transformation dif\/fers from the usual one $\caF$ only by the change of variable $k\rightarrow\pm\widetilde k$.

It is then possible to show the following properties:
\begin{gather}
\int\dd^Dx\,\phi^{2}(x) =\int\dd^Dk\,(\hat\phi\star\hat\phi)(k),\nonumber\\
\int\dd^Dx\,(\phi\star\phi\star\phi\star\phi)(x) =\int\dd^Dk\,(\hat\phi\star\hat\phi\star\hat\phi\star\hat\phi)(k),\label{eq-inter-propfour}\\
\widehat{\partial_\mu\phi}(k_a)=-i(-1)^a(\widetilde k_a)_\mu\hat\phi(k_a),
 \qquad\widehat{(\wx_\mu\phi)}(k_a)=i(-1)^a\partial_\mu^k\hat\phi(k_a).\nonumber
\end{gather}
Note that there is a dif\/ference of sign in comparison with the usual Parseval--Plancherel equality $\int\dd^Dx\,\phi^{2}(x)=\int\dd^Dk\,\caF(\phi)(-k)\caF(\phi)(k)$ due to the cyclic convention of sign in~\eqref{eq-inter-symplfour}. Upon using these identities, one f\/inds that
\begin{gather*}
\int\dd^Dx\left(\frac 12(\partial^x_\mu\phi)^2 +\frac{\Omega^2}{2}\wx^2\phi^2 +\frac{m^2}{2}\phi^2\right)=\int\dd^Dk\left(\frac 12\widetilde k^2\hat\phi^2 +\frac{\Omega^2}{2}(\partial_\mu^k\hat\phi)^2 +\frac{m^2}{2}\hat\phi^2\right),
\end{gather*}
so that the action \eqref{eq-qft-actharm} satisf\/ies
\begin{gather}
S[\phi;m,\lambda,\Omega]=\Omega^2  S\left[\hat\phi,\frac{m}{\Omega},\frac{\lambda}{\Omega^2},\frac{1}{\Omega}\right],\label{eq-inter-lscov}
\end{gather}
called the Langmann--Szabo covariance. At the special point $\Omega=1$, the action is invariant under this duality. This symmetry, which does not apply in the case of \eqref{eq-qft-actinit} without the harmonic term, seems to play a crucial role in spoiling the UV/IR mixing. For an adaptation of this duality to the Minkowskian framework, see \cite{Fischer:2010zg}.

However, the Langmann--Szabo duality is not compatible with gauge symmetry, and more generally not well adapted for actions with cubic terms in the f\/ields. The gauge theory \eqref{eq-qft-acteff} is indeed not covariant under this duality, and it is one major problem of this interpretation of the harmonic term.

In \cite{Bieliavsky:2008qy}, the Langmann--Szabo duality and the quadratic operator $(-\partial^2+\Omega^2\wx^2)$ involved in~\eqref{eq-qft-actharm} have been reinterpreted in terms of the classical metaplectic representation constructed from the Heisenberg group. Let us describe this representation, but in dif\/ferent conventions of those used in \cite{Bieliavsky:2008qy,Folland:1989}, for the link with the superalgebraic interpretation (see Subsection~\ref{subsec-disc-grad}).

Firstly, consider the phase space $\gR^{2D}$, whose coordinates are positions $x_\mu$ and impulsions $p_\mu$, and denote by $\omega$ the symplectic structure on $\gR^{2D}$ def\/ined by: for $(x,p),(y,q)\in\gR^{2D}$,
\begin{gather*}
\omega((x,p),(y,q))=x_\mu \Sigma_{\mu\nu}q_\nu+p_\mu\Sigma_{\mu\nu} y_\nu,
\end{gather*}
so that the matrix representing $\omega$ is $\begin{pmatrix} 0 & \Sigma \\ \Sigma & 0 \end{pmatrix}$, where $\Sigma$ is given by \eqref{eq-qft-sigma}. The Heisenberg algebra is the $\gR$-Lie algebra $\kh_D=\gR^{2D}\oplus\gR$, with the following relations: $\forall \, (x,p,s),(y,q,t)\in\kh_D$,
\begin{gather}
[(x,p,s),(y,q,t)]=(0,0,\omega((x,p),(y,q))).\label{eq-inter-heisalg}
\end{gather}
The Heisenberg group $H_D$ is homeomorphic to $\kh_D$, and the exponential map $\exp:\kh_D\to H_D$ is the identity in this identif\/ication. Then, the group law of $H_D$ is given by:
\begin{gather*}
(x,p,s)\fois(y,q,t)=\left(x+y,p+q,s+t+\frac 12\omega((x,p),(y,q))\right),
\end{gather*}
due to the Baker--Campbell--Hausdorf\/f formula. By the Stone--von~Neumann theorem, $H_D$ admits one unique irreducible unitary representation $\rho$ (up to equivalence), such that $\rho(0,0,s)=e^{-is}\gone$. $\rho$ is called the Schr\"odinger representation.

Let us introduce the operators $X_\mu$ and $P_\mu$ on Schwartz functions $f\in\caS(\gR^D)$: $\forall\, y\in\gR^D$,
\begin{gather}
(X_\mu f)(y)=y_\mu f(y),\qquad (P_\mu f)(y)=i\Sigma_{\mu\nu}\frac{\partial f}{\partial y_\nu}(y)\label{eq-inter-weyl}
\end{gather}
satisfying: $[X_\mu,P_\nu]=i\Sigma_{\mu\nu}$. At the level of the Lie algebra $\kh_D$, the inf\/initesimal representation $\dd\rho:\kh_D\to\caB(\caS(\gR^D))$ takes the usual form: $\forall \, (x,p,s)\in\kh_D$, $\forall \, f\in\caS(\gR^D)$, $\forall\, y\in\gR^D$,
\begin{gather*}
(\dd\rho(x,p,s)f)(y)=-i(x_\mu X_\mu+p_\mu P_\mu+s)f(y).
\end{gather*}
By exponentiation, one obtains the expression of $\rho$, representation of $H_D$ on the Hilbert space $L^2(\gR^D)$:
\begin{gather*}
\rho(x,p,s)f(y)=e^{\frac i2x\Sigma p-ixy-is}f(y-\Sigma p).
\end{gather*}

The symplectic group $Sp(\gR^{2D},\omega)$, def\/ined by automorphisms $M$ of $\gR^{2D}$ satisfying $M^T\omega M=\omega$ in matrix notations, acts naturally on the Heisenberg algebra. A matrix $M=\begin{pmatrix} A & B \\ C & D \end{pmatrix}$ belongs to this group if and only if
\begin{gather*}
C^T\Sigma A+A^T\Sigma C=0, \qquad D^T\Sigma B+B^T\Sigma D=0, \qquad D^T\Sigma A+B^T\Sigma C= \Sigma.
\end{gather*}
The Lie algebra of this group is denoted by $\mathfrak{sp}(\gR^{2D},\omega)$ and contains matrices $\begin{pmatrix} A & B \\ C & D \end{pmatrix}$ such that:
\begin{gather*}
(\Sigma A)^T=\Sigma D,\qquad (\Sigma B)^T=\Sigma B,\qquad (\Sigma C)^T=\Sigma C.
\end{gather*}

Let us start the construction of the metaplectic representation. If $M\in Sp(\gR^{2D},\omega)$, consider the operator $\rho\circ M(x,p,s)=\rho(M.(x,p),s)$ for $(x,p,s)\in H_D$. $\rho\circ M$ is a unitary representation of $H_D$ on the Hilbert space $L^2(\gR^D)$ such that $\rho\circ M(0,0,s)=e^{-is}\gone$. Consequently, $\rho\circ M$ and~$\rho$ are equivalent: there exists a unitary operator $\mu(M)$ on $L^2(\gR^D)$ satisfying: $\forall\,  (x,p,s)\in H_D$,
\begin{gather*}
\rho\circ M(x,p,s)=\mu(M)\rho(x,p,s)\mu(M)^{-1}.
\end{gather*}
$\mu:Sp(\gR^{2D},\omega)\to\caL(L^2(\gR^D))$ is then a unitary projective representation (or a group representation of the double covering of $Sp(\gR^{2D},\omega)$).

For $w\in\gR^{2D}$ and $N=\begin{pmatrix} A & B \\ C & D \end{pmatrix}\in\mathfrak{sp}(\gR^{2D},\omega)$, we denote
\begin{gather*}
\gP_N(w)=-\frac 12 wN\omega w=-\frac 12 w\begin{pmatrix} B\Sigma & A\Sigma \\ D\Sigma & C\Sigma \end{pmatrix}w.
\\
\gP_N(X,P)=-\frac 12XB\Sigma X-\frac 12 PC\Sigma P-XA\Sigma P+\frac i2\tr(A),
\end{gather*}
if $X=(X_\mu)$ and $P=(P_\mu)$ are the $D$-dimensional vectors of the operators on $\caS(\gR^D)$ def\/ined in~\eqref{eq-inter-weyl}. Then, the inf\/initesimal representation $\dd\mu:\mathfrak{sp}(\gR^{2D},\omega)\to\caL(\caS(\gR^D))$ can be expressed~\cite{Folland:1989} as: $\forall \, f\in\caS(\gR^D)$, $\forall\,  y\in\gR^D$,
\begin{gather*}
(\dd\mu(N)f)(y)=-i\gP_Nf(y)=\left(\frac i2yB\Sigma y-\frac i2\frac{\partial}{\partial y}\Sigma C\frac{\partial}{\partial y}+yA\frac{\partial}{\partial y}+\frac 12\tr(A)\right)f(y).
\end{gather*}
Since $\mu(e^N)=e^{\dd\mu(N)}$, we deduce the expression of the metaplectic representation (up to a sign): $\forall\,  f\in L^2(\gR^D)$,
\begin{alignat}{3}
&\text{if }M=\begin{pmatrix} A & 0 \\ 0 & D \end{pmatrix}\in Sp\big(\gR^{2D},\omega\big), \qquad && \mu(M)f(y) = |\det(A)|^{\frac 12}f\big(A^Ty\big), & \nonumber\\
&\text{if }M=\begin{pmatrix} \gone & B \\ 0 & \gone \end{pmatrix}\in Sp\big(\gR^{2D},\omega\big),\qquad & & \mu(M)f(y) = e^{\frac i2 yB\Sigma y}f(y),& \label{eq-inter-meta}\\
&\text{if }M=\begin{pmatrix} 0 & \Sigma \\ \Sigma & 0 \end{pmatrix}\in Sp\big(\gR^{2D},\omega\big),\qquad && \mu(M)f(y) =i^{\frac D2}\caF^{-1}f(y),& \nonumber
\end{alignat}
where the (usual) Fourier transformation is given by
\begin{gather*}
\caF f(y)=\frac{1}{(2\pi)^{\frac D2}}\int\dd^Dz\, f(z)e^{iyz}
\end{gather*}
and the group $Sp(\gR^{2D},\omega)$ is generated by the above matrices.

Before reformulating the interpretation of the Langmann--Szabo duality with the metaplectic representation, we prove the following classical lemma.
\begin{lemma}
\label{lem-inter-ad}
For $M\in Sp(\gR^{2D},\omega)$ and $N\in \mathfrak{sp}(\gR^{2D},\omega)$, one has:
\begin{gather*}
\dd\mu(\Ad_M N)=\mu(M)(\dd\mu(N))\mu(M)^{-1}.
\end{gather*}
\end{lemma}
\begin{proof}
Indeed,
\begin{gather*}
\dd\mu(\Ad_M N)=\frac{\dd}{\dd t}\mu\big(e^{t\Ad_M N}\big)_{|t=0}= \frac{\dd}{\dd t}\mu\big(\Ad_M e^{tN}\big)_{|t=0}=\mu(M)(\dd\mu(N))\mu(M)^{-1}.\tag*{\qed}
\end{gather*}\renewcommand{\qed}{}
\end{proof}

The quadratic operator involved in \eqref{eq-qft-actharm} can be obtained from the inf\/initesimal metaplectic representation as follows:
\begin{gather*}
\text{if }Z_\Omega=\begin{pmatrix} 0 & \frac{4\Omega^2}{\theta^2}\Sigma \\ \Sigma & 0 \end{pmatrix},\qquad \dd\mu(Z_\Omega)=-\frac i2\big(-\partial^2+\Omega^2\widetilde y^2\big).
\end{gather*}
Let us now investigate the elements $M=\begin{pmatrix} A & B \\ C & D \end{pmatrix}\in Sp(\gR^{2D},\omega)$ which leave $Z_\Omega$ invariant or covariant.
\begin{definition}
A matrix $M$ leaves $Z_\Omega$ covariant if $\Ad_M Z_\Omega$ is of the form of $Z_\Omega$ itself, namely
\begin{gather*}
\Ad_M Z_\Omega=\begin{pmatrix} 0  & \alpha\Sigma \\ \beta\Sigma & 0\end{pmatrix},
\end{gather*}
for $\alpha,\beta\in\gR$. This is a reformulation of the Langmann--Szabo covariance \eqref{eq-inter-lscov}.
\end{definition}
Then, by applying Lemma \ref{lem-inter-ad}, the operators $\mu(M)$ applied on the f\/ield $\phi$ will then leave the quadratic part of the theory \eqref{eq-qft-actharm} invariant or covariant. The elements $M$ of $Sp(\gR^{2D},\omega)$ leaving~$Z_\Omega$ covariant have to satisfy:
\begin{gather*}
BD^T+\frac{4\Omega^2}{\theta^2}AC^T=0,\qquad BB^T+\frac{4\Omega^2}{\theta^2}AA^T=\alpha\gone,\qquad DD^T+\frac{4\Omega^2}{\theta^2}CC^T=\beta\gone.
\end{gather*}
Consider now the two opposite typical cases.
\begin{itemize}\itemsep=0pt
\item If $B=C=0$, one obtains the group $O(\gR^{D},\gone)\cap Sp(\gR^{D},\Sigma)$, which was already known to let the theory \eqref{eq-qft-actharm} invariant (see \cite{deGoursac:2009fm}).
\item If $A=D=0$, for instance $M=M_{\text{LS}}=\frac{\theta}{2}\begin{pmatrix} 0 & \frac{4}{\theta^2}\gone \\ \gone & 0 \end{pmatrix}$, then $\Ad_M Z_\Omega=\begin{pmatrix} 0 & \frac{4}{\theta^2}\Sigma \\ \Omega^2\Sigma & 0 \end{pmatrix}$. Since $M_{\text{LS}}=\begin{pmatrix} -\frac{2}{\theta}\Sigma & 0 \\ 0 & -\frac{\theta}{2}\Sigma \end{pmatrix} \omega$, one can compute the operator $\mu(M_{\text{LS}})$ thanks to equations \eqref{eq-inter-meta}: $\forall \, f\in L^2(\gR^D)$,
\begin{gather*}
\mu(M_{\text{LS}})f(y)=\frac{i^{\frac D2}}{(\pi\theta)^{\frac D2}}\int\dd^Dz\,f(z)e^{-iy\wedge z}= i^{\frac D2}\hat f(y),
\end{gather*}
that is the symplectic Fourier transformation \eqref{eq-inter-symplfour} (with even $a$).
\end{itemize}
The Langmann--Szabo duality, at the quadratic level of the action \eqref{eq-qft-actharm}, corresponds then to the adjoint action of $M_{\text{LS}}$ on $Z_\Omega$ (see Lemma \ref{lem-inter-ad}), and it is a symmetry of the theory because $M_{\text{LS}}$ leaves $Z_\Omega$ covariant.

Like at the quadratic level of the action, we consider the elements $M=\begin{pmatrix} A & B \\ C & D \end{pmatrix}\!\in\! Sp(\gR^{2D},\omega)$ leaving the interaction term $\int\dd^4x\ \phi\star\phi\star\phi\star\phi(x)$ invariant under the action $\phi\mapsto \mu(M)\phi$. By using equations \eqref{eq-inter-meta}, we f\/ind the group generated on the one hand by the matrices $\begin{pmatrix} A & 0 \\ 0 & -\Sigma(A^T)^{-1}\Sigma \end{pmatrix}$, where $A\in Sp(\gR^D,\Sigma)$, which correspond to the usual symplectic group of the position space and was already known to let this term invariant; and on the other hand by the matrix $M=M_{\text{LS}}$, def\/ined above. The Langmann--Szabo invariance of the quartic term is indeed due to pro\-per\-ty~\eqref{eq-inter-propfour}.

Thus, we see that the action without harmonic term \eqref{eq-qft-actinit} respects the $Sp(\gR^D,\Sigma)$ symmetry of the quartic term, but not the $M_{\text{LS}}$-symmetry (it is not Langmann--Szabo covariant). To include this symmetry, one can take $\dd\mu(Z_\Omega)$ as the quadratic operator of the action (like in \eqref{eq-qft-actharm}). But this action breaks now the $Sp(\gR^D,\Sigma)$ part into $O(\gR^{D},\gone)\cap Sp(\gR^{D},\Sigma)$. The above reinterpretation is in fact a group explanation of the Langmann--Szabo duality via the metaplectic representation~$\mu$.

\subsection{Superalgebraic framework}
\label{subsec-inter-super}

The second mathematical interpretation of the harmonic term has been given in \cite{deGoursac:2008bd} by introducing a $\gZ_2$-graded algebra and a noncommutative dif\/ferential calculus based on its derivations.

Upon using basic properties of the Moyal product such as $i[\xi_\mu,f]_\star=\partial_\mu f$ and $\{\xi_\mu,f\}_\star=2\xi_\mu f$, where we recall that $\xi_\mu=-\frac 12\wx_\mu$, the action \eqref{eq-qft-actharm} can be reexpressed as:
\begin{gather*}
S(\phi)=\int \dd^Dx\left(\frac 12(i[\xi_\mu,\phi]_\star)^2 +\frac{\Omega^2}{2}(\{\xi_\mu,\phi\}_\star)^2 +\frac{m^2}{2}\phi^2 +\lambda\phi\star\phi\star\phi\star\phi\right).
\end{gather*}
And, at the quadratic level of the action, the Langmann--Szabo duality, which exchanges~$p_\mu$ and~$\wx_\mu$, can be viewed as a symmetry between commutators $i[\fois,\fois]_\star$ and anticommutators $\{\fois,\fois\}_\star$.

Moreover, by replacing the curvature $F_{\mu\nu}=\Theta^{-1}_{\mu\nu}-i[\caA_\mu,\caA_\nu]_\star^2$ in function of the covariant coordinate $\caA_\mu$ in the gauge-invariant action \eqref{eq-qft-acteff}, one f\/inds:
\begin{gather*}
S(\caA)=\int\dd^Dx\left(\frac 14(i[\caA_\mu,\caA_\nu]_\star)^2+\frac{\Omega^2}{4}(\{\caA_\mu,\caA_\nu\}_\star)^2+\kappa \caA_\mu\caA_\mu\right)
\end{gather*}
up to a constant term. For this action, the Langmann--Szabo duality does not apply but it is also symmetric under the exchange of commutators and anticommutators like in the scalar case. This symmetry, which generalizes the Langmann--Szabo duality to the gauge case, seems to be crucial for noncommutative QFT with harmonic term.

The exchange of commutators and anticommutators is very reminiscent to a grading symmetry, and this motivates the introduction of the $\gZ_2$-graded associative algebra $\algrA_\theta=\algA_\theta^0\oplus\algA_\theta^1$, where $\algA_\theta^0=\caM$ and $\algA_\theta^1=\caM$, with product: $\forall\,  f=(f_0,f_1),g=(g_0,g_1)\in\algrA_\theta$,
\begin{gather*}
f\fois g=(f_0\star g_0+\alpha\ f_1\star g_1,f_0\star g_1+f_1\star g_0),
\end{gather*}
where $\alpha$ is a real parameter. The algebra is in fact a quadratic extension of the ring $\caM$, but endowed with a canonical grading compatible with the product. This algebra will be called in the following the Moyal superalgebra.

The graded center of $\algrA_\theta$, namely the elements of $\algrA_\theta$ which (graded-) commute with all elements, is trivial: $\caZ^\bullet(\algA_\theta)=\gC\oplus\algzero=\gC\gone$, where $\gone=(1,0)$ is the unit of the product. One can choose $f^\ast=(f_0^\dag,f_1^\dag)$ as involution on $\algrA_\theta$ and $\tr(f)=\int \dd^Dx\ f_0(x)$ as (non-graded) trace.

The graded bracket associated to $\algrA_\theta$ is given by the following expression: $\forall\,  f,g\in\algrA_\theta$,
\begin{gather*}
[f,g]=([f_0,g_0]_\star+\alpha\{f_1,g_1\}_\star,[f_0,g_1]_\star+[f_1,g_0]_\star)
\end{gather*}
in terms of the commutator and the anticommutator of $\caM$, which will permit to encode the symmetry noticed above.

Then, for $\kg^\bullet$ a graded Lie subalgebra of the graded derivations\footnote{The following def\/inition of derivations applies rigorously only for homogeneous elements $\kX$ and $f$, because the degrees $|\kX|,|f|\in\gZ_2$ are considered. Let us take as a convention that it means the sum of homogeneous components thanks to the linearity.} of $\algrA_\theta$:
\begin{gather*}
\Der^\bullet(\algA_\theta)=\{\kX\in\caL(\algA_\theta),\quad\forall\, f,g\in\algrA_\theta,\qquad \kX(f\fois g)=\kX(f)\fois g+(-1)^{|\kX||f|}f\fois\kX(g)\},
\end{gather*}
one can construct the derivation-based dif\/ferential calculus $\Omega^{\bullet,\bullet}(\algA_\theta|\kg)$ restricted to the derivations involved in $\kg^\bullet$. A $n$-form $\omega\in\Omega^{n,\bullet}(\algA_\theta|\kg)$ is a n-linear map $\omega:(\kg^\bullet)^n\to\algrA_\theta$ with the property of graded antisymmetry: $\forall\, \kX_j\in\kg^\bullet$,
\begin{gather*}
\omega(\kX_1,\dots,\kX_i,\kX_{i+1},\dots,\kX_n)=-(-1)^{|\kX_i||\kX_{i+1}|}\omega(\kX_1,\dots,\kX_{i+1},\kX_i,\dots,\kX_n).
\end{gather*}
For more details on the construction and properties of this noncommutative dif\/ferential calculus, see \cite{DuboisViolette:1988cr,Masson:2008uq} in the case of associative algebras, and \cite{deGoursac:2008bd} for graded associative algebras.

Let us now choose the graded Lie subalgebra $\kg^\bullet$ of derivations. We introduce $\tilde\kg^\bullet$ as generated by the following elements $(0,i)$, $(i\xi_\mu,0)$, $(0,i\xi_\mu)$ and $(i\eta_{\mu\nu},0)$, where $\eta_{\mu\nu}=2\xi_\mu\xi_\nu=\frac 12\wx_\mu \wx_\nu$, which satisfy the relations\footnote{The missing relations vanish.}:
\begin{gather}
[(0,i),(0,i)]=-2\alpha\gone,\qquad
[(0,i\xi_\mu),(0,i)]=(-2\alpha\xi_\mu,0),\qquad
[(i\xi_\mu,0),(i\xi_\nu,0)]=i\Theta^{-1}_{\mu\nu}\gone,\nonumber\\
[(i\xi_\mu,0),(0,i\xi_\nu)]=(0,i\Theta^{-1}_{\mu\nu}),\qquad
[(0,i\xi_\mu),(0,i\xi_\nu)]=(-\alpha\eta_{\mu\nu},0),\nonumber\\
[(i\eta_{\mu\nu},0),(i\xi_\rho,0)]=\left(\frac i2\xi_\mu\Theta^{-1}_{\nu\rho}+\frac i2\xi_\nu\Theta^{-1}_{\mu\rho},0\right),\label{eq-inter-grlie}\\
[(i\eta_{\mu\nu},0),(0,i\xi_\rho)]=\left(0,\frac i2\xi_\mu\Theta^{-1}_{\nu\rho}+\frac i2\xi_\nu\Theta^{-1}_{\mu\rho}\right),\nonumber\\
[(i\eta_{\mu\nu},0),(i\eta_{\rho\sigma},0)]=\left(\frac i2\eta_{\mu\rho}\Theta^{-1}_{\nu\sigma} +\frac i2\eta_{\mu\sigma}\Theta^{-1}_{\nu\rho} +\frac i2\eta_{\nu\rho}\Theta^{-1}_{\mu\sigma} +\frac i2\eta_{\nu\sigma}\Theta^{-1}_{\mu\rho},0\right).\nonumber
\end{gather}
Then, we take $\kg^\bullet=\ad\tilde\kg^\bullet$. Since the (graded) adjoint representation evaluated on $f\in\algrA_\theta$ is a~graded derivation: $\ad_f g=[f,g]$, and because of relations \eqref{eq-inter-grlie}, $\kg^\bullet$ is a graded Lie algebra.

The dif\/ferential acts on the elements of $\algrA_\theta$ as:
\begin{gather*}
 \dd f(\ad_{(i\xi_\mu,0)})=(\partial_\mu f_0,\partial_\mu f_1),\qquad \dd f(\ad_{(0,i\xi_\mu)})=(i\alpha\wx_\mu f_1,\partial_\mu f_0),\\
 \dd f(\ad_{(0,i)})=(-2if_1,0),\qquad \dd f(\ad_{(i\eta_{\mu\nu},0)})=-\frac 14(\wx_\mu\partial_\nu f_0+\wx_\nu\partial_\mu f_0,\wx_\mu\partial_\nu f_1+\wx_\nu\partial_\mu f_1).
\end{gather*}
Because of the $\gZ_2$-grading, the space $\Omega^{\bullet,\bullet}(\algA_\theta|\kg)$ is inf\/inite-dimensional, and one def\/ines neither a Hodge operation for noncommutative metrics in this case, nor the related notion of Laplacian. However, it is still possible to def\/ine a scalar action,
\begin{gather}
S=\tr\left(\sum_a|\ad_a(\phi,\phi)|^2\right),\label{eq-inter-scalact}
\end{gather}
mimicking the ordinary case. In \eqref{eq-inter-scalact}, we have imposed that $\phi_0=\phi_1=\phi\in\caM$ and the sum on $a$ is over $\big\{\big(0,\frac{i}{\sqrt{\theta}}\big),(i\xi_\mu,0),(0,i\xi_\mu),(i\sqrt{\theta}\eta_{\mu\nu},0)\big\}$, where $\sqrt\theta$ appears because of dimensional reasons. Then, we f\/ind
\begin{gather*}
S(\phi)=\int d^Dx\left((1+2\alpha)(\partial_\mu\phi)^2+\alpha^2(\wx_\mu\phi)^2+\frac{4\alpha^2}{\theta}\phi^2\right)
\end{gather*}
as a part of the action \eqref{eq-inter-scalact}, so that the Grosse--Wulkenhaar model \eqref{eq-qft-actharm} emerges from this superalgebraic framework. Moreover, the Langmann--Szabo duality is related to the grading symmetry $(i\xi_\mu,0)\rightleftarrows(0,i\xi_\mu)$ at the quadratic level of the action~\eqref{eq-qft-actharm} (see Subsection~\ref{subsec-disc-grad}).

In this setting, a graded connection on $\algrA_\theta$ (considered as a right module on itself) is a~homogeneous linear map of degree 0, $\nabla:\algrA_\theta\to\Omega^{1,\bullet}(\algA_\theta|\kg)$ such that: $\forall\, \kX\in\kg^\bullet$, $\forall f,g\in\algrA_\theta$,
\begin{gather*}
\nabla(f\fois g)(\kX)=f\fois\dd g(\kX)+(-1)^{|g||\kX|}\nabla(f)(\kX)\fois g.
\end{gather*}
Let us denote by $A_\mu^0$, $A_\mu^1$, $\varphi$ and $G_{\mu\nu}$ the gauge potentials associated to $\nabla$:
\begin{alignat*}{3}
& \nabla(\gone)(\ad_{(i\xi_\mu,0)})=(-iA_\mu^0,0),\qquad &&
\nabla(\gone)(\ad_{(0,i\xi_\mu)})=(0,-iA_\mu^1),&\\
& \nabla(\gone)(\ad_{(0,i)})=(0,-i\varphi),\qquad &&
\nabla(\gone)(\ad_{(i\eta_{\mu\nu},0)})=(-iG_{\mu\nu},0).&
\end{alignat*}
The gauge transformations are then dictated by the theory. They are determined by unitary elements $g$ of $\caM$ ($g^\dag\star g=g\star g^\dag=1$) acting on the f\/ields as:
\begin{gather*}
\big(A_\mu^0\big)^g=g\star A_\mu^0\star g^\dag+ig\star\partial_\mu g^\dag, \qquad
\big(A_\mu^1\big)^g=g\star A_\mu^1\star g^\dag+ig\star\partial_\mu g^\dag,\\
\varphi^g=g\star\varphi\star g^\dag,\qquad
(G_{\mu\nu})^g=g\star G_{\mu\nu}\star g^\dag-\frac i4g\star(\wx_\mu\partial_\nu g^\dag)-\frac i4g\star(\wx_\nu\partial_\mu g^\dag).
\end{gather*}

Then, the graded curvature $F_{\kX,\kY}$, for $\kX,\kY\in\kg^\bullet$, can be computed:
\begin{gather*}
F_{\kX,\kY}=\kX(A_\kY)-(-1)^{|\kX|,|\kY|}\kY(A_\kX)-i[A_\kX,A_\kY]-A_{[\kX,\kY]}
\end{gather*}
in term of the potentials $A_\kX=i\nabla(\gone)(\kX)$, and it turns out that the Yang--Mills action\footnote{With an implicit summation on $a,b$ like in the scalar case \eqref{eq-inter-scalact} above.}\linebreak  $\tr(|F_{\ad_a,\ad_b}|^2)$ for this model takes the following form\footnote{By identifying the f\/ields $A_\mu^0=A_\mu^1=A_\mu$, which have the same gauge transformations.}:
\begin{gather}
S(A,\varphi,G)=\int d^Dx\Bigg((1+2\alpha)F_{\mu\nu}\star F_{\mu\nu}+\alpha^2\{\caA_\mu,\caA_\nu\}_\star^2 \nonumber\\
\phantom{S(A,\varphi,G)=}{}
+\frac{8}{\theta}(2(D+1)(1+\alpha)+\alpha^2)\caA_\mu\star\caA_\mu
+2\alpha(\partial_\mu\varphi-i[A_\mu,\varphi]_\star)^2
\nonumber\\
\phantom{S(A,\varphi,G)=}{}
+2\alpha^2(\wx_\mu\varphi+\{A_\mu,\varphi\}_\star)^2-4\alpha\sqrt{\theta}\varphi\Theta_{\mu\nu}^{-1}F_{\mu\nu}
+\frac{2\alpha(D+2\alpha)}{\theta}\varphi^2-\frac{8\alpha^2}{\sqrt{\theta}}\varphi\star\varphi\star\varphi\nonumber\\
\phantom{S(A,\varphi,G)=}{}
 +4\alpha^2\varphi\star\varphi\star\varphi\star\varphi -\alpha[\caG_{\mu\nu},\varphi]_\star^2 +2\alpha^2\{\caA_\mu,\caA_\nu\}_\star\star\caG_{\mu\nu}\nonumber\\
\phantom{S(A,\varphi,G)=}{}
-8i(1+\alpha)\Theta^{-1}_{\nu\mu}[\caA_\rho,\caA_\mu]_\star\star\caG_{\nu\rho} +8i\caG_{\mu\nu}\star\caG_{\rho\sigma}\star(\Theta^{-1}_{\mu\rho}\caG_{\nu\sigma} +\Theta^{-1}_{\nu\sigma}\caG_{\mu\rho} +\Theta^{-1}_{\nu\rho}\caG_{\mu\sigma}\nonumber\\
\phantom{S(A,\varphi,G)=}{}
 +\Theta^{-1}_{\mu\sigma}\caG_{\nu\rho})
+\bigg(\frac{16}{\theta^2}(D+2)\alpha^2\bigg)\caG_{\mu\nu}\star\caG_{\mu\nu} +2\caG_{\mu\nu}\star\caG_{\mu\nu}\star\caG_{\rho\sigma}\star\caG_{\rho\sigma} \nonumber\\
\phantom{S(A,\varphi,G)=}{}
-2\caG_{\mu\nu}\star\caG_{\rho\sigma}\star\caG_{\mu\nu}\star\caG_{\rho\sigma}\Bigg),\label{eq-inter-actymh}
\end{gather}
where $\caA_\mu=A_\mu+\frac 12\wx_\mu$ and $\caG_{\mu\nu}=G_{\mu\nu}-\frac 12\wx_\mu\wx_\nu$.

The pure part of the action \eqref{eq-inter-actymh} in the variable $A_\mu$ (or $\caA_\mu$) gives the action \eqref{eq-qft-acteff}, while the pure part in the variable $\varphi$ is a scalar action, with $\star$-polynomial and harmonic term, of the type~\eqref{eq-qft-actharm}. The coupling between the two f\/ields is the standard one \cite{deGoursac:2007gq} for a gauge f\/ield $A_\mu$ and a scalar f\/ield $\varphi$ in the ``adjoint representation'' of the gauge group, but there is also an additional BF-term ($-4\alpha\sqrt\theta\varphi\Theta^{-1}_{\mu\nu}F_{\mu\nu}$). Therefore, both Grosse--Wulkenhaar model and its associated gauge theory can be obtained in this graded formalism by computing a graded curvature for the noncommutative derivation-based dif\/ferential calculus of the Moyal superalgebra $\algrA_\theta$. For more details on the calculations, see \cite{deGoursac:2008bd}. Notice that in the dif\/ferent context of commutative theories, the quadratic operator with harmonic term has been obtained in~\cite{Wulkenhaar:2009pv} by considering also a $\gZ_2$-graded Lie algebra of operators acting on a certain Hilbert space.

We will see in Subsection \ref{subsec-disc-grad} that this approach leading to the action \eqref{eq-inter-actymh} reproduces the Langmann--Szabo duality by a grading symmetry, and it is well adapted to gauge theory, contrary to the Langmann--Szabo duality itself. However, this theory involves two supplementary f\/ields: one ($\varphi$) can be interpreted as a Higgs f\/ield\footnote{Note that such a Higgs f\/ield has already been obtained by a similar (but non-graded) procedure on the Moyal space in~\cite{Cagnache:2008tz}.}, but the interpretation of the other one ($G_{\mu\nu}$) is less obvious. In \cite{deGoursac:2008bd}, it has been suggested that it could play the role of the symmetric counterpart of a BF-f\/ield.

\subsection{Noncommutative scalar curvature}
\label{subsec-inter-scalcurv}

In this subsection, we will describe how to obtain the harmonic term $x^2$ of the action \eqref{eq-qft-actharm} as a noncommutative scalar curvature \cite{Buric:2009ss}. Indeed, the Moyal algebra can be approximated by matrix algebras, just as the Heisenberg algebra is a limit of ``truncated Heisenberg algebras''. These are (non-graded) quadratic algebras, on which a noncommutative dif\/ferential calculus has been constructed in \cite{Buric:2009ss} by using the Cartan frame formalism \cite{Madore:1996bb}. Then, linear connections can be introduced and scalar curvature of metric connections can be computed \cite{Dubois-Violette:1996aa}. By taking the limit of these curvatures, one f\/inds the term $x^2$. We expose here the $D=2$-dimensional case for simplicity reasons.

It has been shown in \cite{GraciaBondia:1987kw} that the eigenfunctions $(f_{mn}(x))$, for $m,n,\in\gN$ in the two-dimen\-sio\-nal case, of the harmonic oscillator $H=\frac{x^2}{2}$ provide a basis of $\caM$ (with respect to its Fr\'echet structure), called the matrix basis. These functions are def\/ined by:
\begin{gather*}
H\star f_{mn}=\theta\left(m+\frac 12\right)f_{mn},\qquad f_{mn}\star H=\theta\left(n+\frac 12\right)f_{mn},
\end{gather*}
and they are Schwartz functions. It turns out that they satisfy the following properties:
\begin{gather}
f_{mn}\star f_{kl}=\delta_{nk}f_{ml},\qquad \int\dd^2x\,f_{mn}(x)=2\pi\theta\delta_{mn}, \qquad f_{mn}^\dag=f_{nm}.\label{eq-inter-propmatrix}
\end{gather}
Each element $g\in\caM$ can then be written as:
\begin{gather*}
g=\sum_{m,n\in\gN}g_{mn}f_{mn},
\end{gather*}
where the coef\/f\/icients $g_{mn}\in\gC$ are given by $g_{mn}=\frac{1}{2\pi\theta}\int\dd^2x\,(g\ f_{nm})$. Because of properties~\eqref{eq-inter-propmatrix}, the coef\/f\/icients $(g_{mn})$ behave like matrices under the product, the trace and the involution. For example, the coef\/f\/icient of the Moyal product of two elements is just the matrix product of the coef\/f\/icients of these two elements: $(g\star h)_{mn}=\sum_{k\in\gN}g_{mk}h_{kn}$. In this identif\/ication, $\caM$ is a~limit of f\/inite-dimensional matrix algebras.

Let us exhibit what could be an approximation of the Heisenberg algebra $\kh_1=\langle 1,x_1,x_2\rangle$ (satisfying $[x_1,x_2]_\star=-i\theta$) in this context. The matrix coef\/f\/icient of the function $x\mapsto 1$ is $\delta_{mn}$, while those of the coordinates are:
\begin{gather*}
(x_1)_{mn}=\sqrt{\frac{\theta}{2}}\big(\sqrt n\delta_{m+1,n}+\sqrt m\delta_{m,n+1}\big),\qquad (x_2)_{mn}=i\sqrt{\frac{\theta}{2}}\big(\sqrt n\delta_{m+1,n}-\sqrt m\delta_{m,n+1}\big).
\end{gather*}
Then, following \cite{Buric:2009ss}, we introduce the $n\times n$ matrices:
\begin{gather*}
X_n=\frac{1}{\sqrt 2}\begin{pmatrix} 0 & 1 & 0 &  & \\ 1 & 0 & \sqrt 2 &  & \\ 0 & \sqrt 2 & 0 & \ddots & \\  &  & \ddots &  & \sqrt{n-1} \\  &  & & \sqrt{n-1} & 0 \end{pmatrix},\\
Y_n=\frac{i}{\sqrt 2}\begin{pmatrix} 0 & -1 & 0 &  & \\ 1 & 0 & -\sqrt 2 &  & \\ 0 & \sqrt 2 & 0 & \ddots & \\  &  & \ddots &  & -\sqrt{n-1} \\  &  & & \sqrt{n-1} & 0 \end{pmatrix},
\end{gather*}
which verify the commutation relation $[X_n,Y_n]=i(\gone-Z_n)$, where $\gone$ is the unit matrix and $Z_n=\begin{pmatrix} 0 & 0 &  & \\ 0 & 0 & \ddots & \\  & \ddots &  & 0 \\  & & 0 & n \end{pmatrix}$. By considering these three generators, one obtains the following quadratic algebra, called the ``truncated Heisenberg algebra'' $\kh^{(n)}_1$:
\begin{gather*}
[X_n,Y_n]=i(\gone-Z_n),\qquad [X_n,Z_n]=i(Y_nZ_n+Z_nY_n),\qquad [Y_n,Z_n]=-i(X_nZ_n+Z_nX_n).
\end{gather*}
When $n\to\infty$, one has: $X_n\to\frac{1}{\sqrt\theta}x_1$, $Y_n\to -\frac{1}{\sqrt\theta}x_2$ and $Z_n\to 0$, so that these quadratic algeb\-ras~$\kh^{(n)}_1$ approximate the Heisenberg algebra $\kh_1$, up to a rescaling of the generators.

The momenta associated to these coordinates are\footnote{We omit the subscript $n$.}: $p_1=iY$, $p_2=-iX$ and $p_3=i(Z-\frac 12)$, satisfying the quadratic relations:
\begin{gather*}
[p_1,p_2]=p_3-\frac i2,\qquad [p_2,p_3]=p_1-i(p_1p_3+p_3p_1),\qquad [p_3,p_1]=p_2-i(p_2p_3+p_3p_2).
\end{gather*}
Note that the quadratic algebra of \cite{Buric:2009ss} contains moreover normalization factors $\epsilon$ and $\mu$ which we do not write here for simpler notations. If we apply the Cartan frame formalism \cite{Madore:1996bb} to this quadratic algebra, we obtain a noncommutative dif\/ferential calculus, whose one-forms are generated by the frame $\{\theta^i\}$, where $\theta^i:\Int(\kh_1^{(n)})\to\gC$ satisf\/ies $\theta^i(\ad_{p_j})=\delta^i_j$. On one-forms, the product and the dif\/ferential are def\/ined by:
\begin{gather*}
(\theta^i)^2=0,\qquad \theta^1\theta^2=-\theta^2\theta^1,\qquad \theta^2\theta^3=-i\theta^1\theta^3,\qquad \theta^3\theta^2=i\theta^3\theta^1,\\
\dd\theta^1=-\left(p_3-\frac i2\right)(\theta^1\theta^3+\theta^3\theta^1),\qquad \dd\theta^2=i\left(p_3-\frac i2\right)(\theta^1\theta^3-\theta^3\theta^1),\\
\dd\theta^3=-\theta^1\theta^2 -p_1 (\theta^1\theta^3+\theta^3\theta^1)+ip_2(\theta^1\theta^3-\theta^3\theta^1).
\end{gather*}

In this setting, one denotes ${\omega^i}_j={\omega^i}_{kj}\theta^k$ the one-form associated to a linear connection. In \cite{Buric:2009ss} is given an approximated solution (in the commutative limit) of a torsion-free linear connection compatible with the (noncommutative) metric $g_{ij}=\delta_{ij}$:
\begin{gather*}
{\omega^1}_2=-{\omega^2}_1=\left(-\frac 12+2ip_3\right)\theta^3,\qquad
{\omega^1}_3=-{\omega^3}_1=\frac 12\theta^2+2ip_2\theta^3,\\
{\omega^2}_3=-{\omega^3}_2=-\frac 12\theta^1-2ip_1\theta^3.
\end{gather*}
Then, one can obtain the noncommutative Riemann curvature of this connection:
\begin{gather*}
{\Omega^i}_j=\dd{\omega^i}_j+{\omega^i}_k{\omega^k}_j=\frac 12 {R^i}_{jkl}\theta^k\theta^l,
\end{gather*}
and its scalar curvature $R=g^{ij}{R^k}_{ikj}$. In this case, one f\/inds \cite{Buric:2009ss}:
\begin{gather*}
R=\frac{11}{2}+4ip_3+8(p_1^2+p_2^2)=\frac{15}{2}-4Z-8\big(X^2+Y^2\big).
\end{gather*}
In the limit $n\to\infty$, the scalar curvature takes the form: $R=\frac{15}{2\theta}-8\wx^2$ up to a rescaling.

On a commutative but curved manifold, the classical action of a scalar theory involves na\-tu\-ral\-ly a term like $\int\sqrt{g}\dd x(\xi R\phi^2)$, where $R$ is the scalar curvature of the (pseudo-) Riemannian manifold (see \cite{Hollands:2002ux} and references therein). This term is indeed automatically generated by quantum corrections, so that the parameter $\xi$ has to be renormalized.

In that situation, the harmonic term $\wx^2$ can be interpreted as a noncommutative scalar curvature, and the Grosse--Wulkenhaar model \eqref{eq-qft-actharm} would be a generalization of curved scalar f\/ield theories to a noncommutative framework. Notice that this harmonic tem is not generated by quantum corrections in the scalar model \eqref{eq-qft-actinit}, contrary to the curved commutative models.

However, this interpretation is not directly possible on the Moyal algebra. Indeed, the Heisenberg algebra $\kh_1$ is a Lie algebra and not a quadratic algebra. Therefore, the application of the Cartan formalism would give in this case a vanishing scalar curvature $R$. One had to approximate this algebra by the ``truncated Heisenberg algebras''.

Recently, the gauge model associated to this noncommutative dif\/ferential calculus has been exhibited in \cite{Buric:2010xs}. As usual, the curvature $\tF$, or f\/ield strength, can be computed in terms of the gauge potential $A$ associated to a gauge connection by the formula: $\tF=\dd A+A^2$, where $A=A_i\theta^i$ and $\tF=\frac 12 \tF_{ij}\theta^i\theta^j$ are respectively a one-form and a two-form. One obtains:
\begin{gather*}
\tF_{12} =[p_1,A_2]-[p_2,A_1]+[A_1,A_2]-A_3,\\
\tF_{13} =[p_1+A_1,A_3]-i\{p_2+A_2,A_3\}+2A_2Z,\\
\tF_{23} =[p_2+A_2,A_3]+i\{p_1+A_1,A_3\}-2A_1Z.
\end{gather*}
Then, we restrict to the case of the Moyal space: $n\to\infty$ (which means $Z\to0$). By considering the gauge transformations of the potential:
\begin{gather*}
A_1\mapsto gA_1 g^{-1}+g\big[p_1,g^{-1}\big],\qquad A_2\mapsto gA_2 g^{-1}+g[p_2,g^{-1}],\qquad A_3\mapsto g A_3 g^{-1},
\end{gather*}
we denote $A_3=\varphi$, considered as a scalar f\/ield in the adjoint representation. The curvature takes the form:
\begin{gather*}
\tF_{12} =[\caA_1,\caA_2]_\star+i-\varphi,\qquad
\tF_{13} =[\caA_1,\varphi]_\star-i\{\caA_2,\varphi\}_\star,\qquad
\tF_{23} =[\caA_2,\varphi]_\star+i\{\caA_1,\varphi\}_\star,
\end{gather*}
where $\caA_\mu$ is the covariant coordinate introduced in Subsection~\ref{subsec-inter-super}. By using a Hodge operation def\/ined in \cite{Buric:2010xs}, the following Yang--Mills action is provided\footnote{This is a reformulation of what has been found in \cite{Buric:2010xs} in our notations. We also have reintroduced the parameters $\epsilon$ and $\mu$.}:
\begin{gather}
S(A,\varphi)=\int\dd^2x\bigg(\frac{(1-\epsilon^2)}{4}F_{\mu\nu}\star F_{\mu\nu} +\frac{(5-\epsilon^2)\mu^2}{2}\varphi^2
+\frac{\mu\theta(1-\epsilon^2)}{2}\varphi\Theta^{-1}_{\mu\nu}F_{\mu\nu}\nonumber\\
\phantom{S(A,\varphi)=}{}
-i\epsilon\theta(\varphi\star\varphi)\Theta^{-1}_{\mu\nu}F_{\mu\nu}
-\frac 12(\partial_\mu\varphi-i[A_\mu,\varphi]_\star)^2-\frac{\epsilon^2}{2}
(\wx_\mu\varphi+\{A_\mu,\varphi\}_\star)^2\bigg),\label{eq-inter-actgaugecartan}
\end{gather}
where $F_{\mu\nu}=\partial_\mu A_\nu-\partial_\nu A_\mu-i[A_\mu,A_\nu]_\star$ is the standard curvature for the Moyal algebra, and $\mu,\nu\in\{1,2\}$.

This action is completed by the following gauge f\/ixing term: $S_{\text{gf}}=\int s(\overline c\caF+\frac{\alpha}{2}\overline cB)$, where $\caF$ is the gauge, $c$ the ghost f\/ield, $\overline c$ the antighost f\/ield, $B$ the auxiliary f\/ield, and the Slavnov operator $s$ is acting as:
\begin{gather*}
sA_\mu=\partial_\mu c-i[A_\mu,c]_\star,\qquad s\varphi=-i[\varphi,c]_\star,\qquad sc=-c^2,\qquad s\overline c=B,\qquad sB=0.
\end{gather*}
Here, the chosen gauge is the covariant one:
\begin{gather*}
\caF=\nabla_i A_i=[p_i,A_i]+A_j{\omega^i}_{ij}=\partial_\nu A_\nu-2\mu^2x_\nu A_\nu,
\end{gather*}
where $\omega$ (and $\nabla$) is the linear connection def\/ined in the f\/irst part of this subsection.

One can observe that the pure part of the action \eqref{eq-inter-actgaugecartan} in the variable $A_\mu$ is of the type~\eqref{eq-qft-actgauge}, while the other terms describe a scalar f\/ield with a harmonic term, coupled to $A_\mu$, and with BF-terms. Except the special term $\int (\varphi\star\varphi)\Theta^{-1}_{\mu\nu}F_{\mu\nu}$, all the terms of \eqref{eq-inter-actgaugecartan} are included in \eqref{eq-inter-actymh}. However, the term $\int\{\caA_\mu,\caA_\nu\}_\star^2$ of \eqref{eq-qft-acteff} is missing here, so that the propagator of the gauge potential is not the Mehler kernel like in \eqref{eq-inter-actymh}, but there is no problem of non-trivial vacuum more. The hope of this solution is then \cite{Buric:2010xs} that the UV/IR mixing of the theory be canceled by the coupling between the gauge potential and a scalar f\/ield whose propagator is the Mehler kernel.

\section{Discussion}
\label{sec-disc}

\subsection[The Langmann-Szabo duality as a grading exchange]{The Langmann--Szabo duality as a grading exchange}
\label{subsec-disc-grad}

We have seen in Subsection \ref{subsec-inter-super} that a symmetry between commutators and anticommutators seems to be fundamental for both scalar theory \eqref{eq-qft-actharm} and gauge theory \eqref{eq-qft-acteff}. In the scalar case, it corresponds at the quadratic level of the action to the Langmann--Szabo duality. Let us show that this duality is exactly the grading exchange. The crucial object which has led us to the group reformulation of the Langmann--Szabo duality in Subsection \ref{subsec-inter-lsd} was the Heisenberg algebra $\kh_D$. In this subsection, we will exhibit the link between $\kh_D$ and the Lie superalgebra of generators $\widetilde\kg^\bullet=\langle(0,i),(i\xi_\mu,0),(0,i\xi_\mu),(i\eta_{\mu\nu},0)\rangle$ of $\kg^\bullet$ (see equation \eqref{eq-inter-grlie}).

Let $\kh_D=\gR^{2D}\oplus\gR$ be the Heisenberg algebra associated to the symplectic form\footnote{The inverse of $\Sigma$, instead of $\Sigma$ itself, is taken only for conventional reasons.} $\omega=\begin{pmatrix} 0 & \Sigma^{-1} \\ \Sigma^{-1} & 0 \end{pmatrix}$ on the phase space $\gR^{2D}$, and whose commutation relations are given by \eqref{eq-inter-heisalg}. Since we want to deform the (symplectic) position space $(\gR^D,\Sigma)$, we consider an {\it extension} $\widetilde\kh_D\simeq \gC^{2D}\oplus\gC\oplus\gC$ of the complexif\/ication of $\kh_D$: $\forall\,  x,y,p,q\in\gC^D$, $\forall \, s,t,a,b\in\gC$,
\begin{gather*}
[(x,p,s,a),(y,q,t,b)]=(0,0,\omega((x,p),(y,q)),i\theta\Sigma^{-1}(x,y)+i\alpha\theta\Sigma^{-1}(p,q)).
\end{gather*}
This is a very standard way to provide noncommutative position coordinates, with deformation parameter $\theta$, and noncommutative impulsion coordinates, with deformation parameter $\alpha\theta$.

Then, one can realize this extension $\widetilde\kh_D$ in $\algrA_\theta$ by the following map: $(x,p,s,a)\mapsto x_\mu\lambda_\mu+ p_\mu\lambda_{\overline\mu} +s\lambda_0+a\gone$, where we set $\lambda_0=(0,i\theta)$, $\lambda_\mu=(i\theta\xi_\mu,0)$ and $\lambda_{\overline\mu}=(0,i\theta\xi_\mu)$ in $\algrA_\theta$. This map is an injective (non-graded) Lie algebra morphism. The non-vanishing commutation relations (with non-graded bracket) can be expressed as:
\begin{gather*}
[\lambda_\mu,\lambda_\nu]=i\theta\Sigma^{-1}_{\mu\nu}\gone,\qquad [\lambda_\mu,\lambda_{\overline\nu}]=\Sigma^{-1}_{\mu\nu} \lambda_0,\qquad [\lambda_{\overline\mu},\lambda_{\overline\nu}]= i\alpha\theta\Sigma^{-1}_{\mu\nu}\gone.
\end{gather*}
In the commutative limit $\theta\to 0$, we f\/ind indeed a Lie algebra isomorphic to $\kh_D$.

The grading of $\algrA_\theta$ brings a natural grading on $\widetilde\kh_D$. The next step of this procedure can be called the {\it ``superization''} of $\widetilde\kh_D$. Indeed, it is a (non-graded, for its bracket) Lie algebra, but now with a $\gZ_2$-grading and included in a graded associative algebra. We can superize the bracket of~$\widetilde\kh_D$ by changing only the bracket of two odd elements into their anticommutator in $\algrA_\theta$. Then, the changed relations hold:
\begin{gather*}
[\lambda_0,\lambda_0]=-2\alpha\theta^2\gone,\qquad [\lambda_{\overline\mu},\lambda_0]=2i\alpha\theta\lambda_\mu, \qquad [\lambda_{\overline\mu},\lambda_{\overline\nu}]= \alpha(\lambda_\mu\lambda_\nu+\lambda_\nu\lambda_\mu)=i\alpha(i\eta_{\mu\nu},0).
\end{gather*}

By rescaling the generators $\lambda_i\mapsto\frac{1}{\theta}\lambda_i$ and taking the {\it closure} in $\algrA_\theta$ (under the graded bracket) of the space $\langle\lambda_i\rangle$, one obtains the Lie superalgebra $\widetilde\kg^\bullet$ described in \eqref{eq-inter-grlie}.

The symplectic group $Sp(\gR^{2D},\omega)$ which was acting by automorphisms on $\kh_D$, is no longer a~symmetry group of $\widetilde\kg^\bullet$. Only the subgroup $Sp(\gR^D,\Sigma)\otimes\begin{pmatrix} 1 & 0 \\ 0 & 1 \end{pmatrix}$ survives to the above procedure. Then, the matrix $M_{\text{LS}}=\frac{\theta}{2}\begin{pmatrix} 0 & \frac{4}{\theta^2}\gone \\ \gone & 0 \end{pmatrix}$ representing the Langmann--Szabo duality at the level of the symplectic group (see Subsection \ref{subsec-inter-lsd}) does not leave the relations~\eqref{eq-inter-grlie} invariant. However, it is still an endomorphism of the graded vector space $\widetilde\kg^\bullet$, and it acts as:
\begin{alignat*}{3}
& M_{\text{LS}}.(i\xi_\mu,0)=\frac{\theta}{2}(0,i\xi_\mu),\qquad && M_{\text{LS}}.(0,i\xi_\mu)=\frac{2}{\theta}(i\xi_\mu,0),&\\
& M_{\text{LS}}.(0,i)=(0,i), \qquad && M_{\text{LS}}.(i\eta_{\mu\nu},0)=(i\eta_{\mu\nu},0),&
\end{alignat*}
which corresponds to the grading exchange for f\/irst order polynomials (up to a rescaling) in $\gR^{2D}$. In Subsection~\ref{subsec-inter-super}, there was only a correspondence between the Langmann--Szabo duality and the grading exchange; we see here by the above procedure that both symmetries are identical. Certainly, the action of $M_{\text{LS}}$ is not a symmetry of $\widetilde\kg^\bullet$. But one restores it as a symmetry of the scalar f\/ield theory by identifying $\phi_0=\phi_1$, and of the gauge theory by identifying $A_\mu^0=A_\mu^1$ (see Subsection~\ref{subsec-inter-super}). Furthermore, the f\/ield $G_{\mu\nu}$ of Subsection~\ref{subsec-inter-super} is auxiliary in this point of view, since it arises only from the closure of the Lie superalgebra.

Let us summarize the discussion of this subsection.
\begin{proposition}
In the above notations, one can obtain the Lie superalgebra $\widetilde\kg^\bullet$ from the Heisenberg algebra $\kh_D$ by the following three-steps procedure:
\begin{enumerate}
\itemsep=0pt
\item Extension, for the deformation quantization,
\item ``Superization'' in $\algrA_\theta$,
\item Closure in $\algrA_\theta$ under the graded bracket.
\end{enumerate}
Moreover, the Langmann--Szabo duality is identical to the grading exchange $(i\xi_\mu,0)\rightleftarrows(0,i\xi_\mu)$.
\end{proposition}

Note that an analogous procedure (but without ``superization'') has been considered in~\cite{Bieliavsky:2008qy} for a rank one Hermitian symmetric space of the non-compact type, instead of $\gR^{2D}$.

\subsection{Relations and comparison between the dif\/ferent viewpoints}
\label{subsec-disc-compar}

We have seen in Subsection~\ref{subsec-inter-lsd} that the Langmann--Szabo duality, which is a cyclic symplectic Fourier transformation, can explain the form of the action \eqref{eq-qft-actharm} of the Grosse--Wulkenhaar model, but it is not well adapted to its associated gauge theory. We have also observed that this duality could be reformulated in terms of the metaplectic representation of the symplectic group, constructed from the Heisenberg algebra $\kh_D$. This algebra contains indeed positions and impulsions, which are exchanged in this duality.

The superalgebraic approach, described in Subsection \ref{subsec-inter-super}, is {\it a priori} a dif\/ferent explanation of the Grosse--Wulkenhaar model. The choice of a graded noncommutative dif\/ferential calculus adapted to a certain Lie superalgebra $\widetilde\kg^\bullet$ gives rise to the scalar action~\eqref{eq-qft-actharm} so as its gauge associated theory \eqref{eq-qft-acteff}. Furthermore, the analysis of Subsection~\ref{subsec-disc-grad} makes appear the unif\/ication of the Langmann--Szabo duality interpretation with the superalgebraic approach. Indeed, the Lie superalgebra $\widetilde\kg^\bullet$ can be obtained by some natural procedure from the Heisenberg algebra $\kh_D$, in the context of the graded algebra $\algrA_\theta$. In this picture, the above noncommutative dif\/ferential calculus is constructed from $\kh_D$, and the Langmann--Szabo duality is in fact the grading exchange for f\/irst order polynomials in $\algrA_\theta$: $(i\xi_\mu,0)\rightleftarrows(0,i\xi_\mu)$.

The gauge action \eqref{eq-inter-actymh}, which contains \eqref{eq-qft-acteff}, has now a geometrical interpretation, as constructed from a graded curvature. Moreover, because of the supplementary f\/ield $G_{\mu\nu}$, this action has the following trivial vacuum: $A_\mu=0$, $\varphi=0$, $G_{\mu\nu}=0$ (see \cite{deGoursac:2008bd}), contrary to the action \eqref{eq-qft-acteff}. And the gauge sector of~\eqref{eq-inter-actymh} (depending only on $A_\mu$) has the Mehler kernel as propagator, up to gauge f\/ixing terms (see equation~\eqref{eq-qft-actquadr}), knowing that this Mehler kernel cures the UV/IR mixing in the scalar case.

In Subsection \ref{subsec-inter-scalcurv}, we have studied another viewpoint of the harmonic term $\widetilde x^2$, obtained as the limit of noncommutative scalar curvature of ``truncated Heisenberg algebras'' $\kh^{(n)}_{\frac D2}$. Indeed, the scalar curvature appears naturally in curved commutative models as generated by quantum corrections. By analogy to this situation, and even if the term $\widetilde x^2$ is not generated by quantum corrections for the action \eqref{eq-qft-actinit}, one can explain this term in the action \eqref{eq-qft-actharm} to be a noncommutative scalar curvature.

The gauge model related to this interpretation has also been described in Subsection~\ref{subsec-inter-scalcurv} (see equation~\eqref{eq-inter-actgaugecartan}), and is quite dif\/ferent from \eqref{eq-qft-acteff}. Since it does not contain the symmetry between commutators and anticommutators, that we have seen in Subsections~\ref{subsec-inter-super} and \ref{subsec-disc-grad} to be a generalization of the Langmann--Szabo duality, one concludes that this viewpoint cannot be unif\/ied with the two f\/irst ones. The hope of renormalizability of the gauge model \eqref{eq-inter-actgaugecartan} is that the coupling between the gauge f\/ield and the scalar f\/ield (with an harmonic term) would cancel the UV/IR mixing. We see that this would be a dif\/ferent mechanism from the gauge model~\eqref{eq-inter-actymh}, where the gauge sector is supposed to be already free of UV/IR mixing.

In fact, the central object of the three mathematical interpretations is the Heisenberg algebra, $\kh_D$ for the two f\/irst ones, and $\kh_{\frac D2}$ (and its truncations) for the last one. Of course, one may have considered the truncated algebras of $\kh_D$, and the resulting scalar curvature would have been proportional to $p^2+\widetilde x^2$, the operator involved in \eqref{eq-qft-actharm}. However, the interpretation by analogy of the curved commutative case is then no longer valid, and the associated gauge theory is still dif\/ferent from \eqref{eq-qft-acteff}. Therefore, there are essentially two dif\/ferent approaches to interpret the noncommutative quantum f\/ield theory with harmonic term, by considering $\kh_D$ or $\kh_{\frac D2}$, namely the Langmann--Szabo duality and the superalgebraic approach, which are now unif\/ied, and the noncommutative scalar curvature interpretation. These both approaches provide non-equivalent gauge models, which are candidates to renormalizability.

\subsection*{Acknowledgements}

The author thanks Pierre Bieliavsky and Jean-Christophe Wallet for interesting discussions at various stages of this work. This work was supported by the Belgian Interuniversity Attraction Pole (IAP) within the framework ``Nonlinear systems, stochastic processes, and statistical mechanics'' (NOSY).

\pdfbookmark[1]{References}{ref}
\LastPageEnding

\end{document}